\documentclass[aps,pre,superscriptaddress,twocolumn,showpacs,floatfix]{revtex4}
\usepackage{graphicx}
\usepackage{dcolumn}
\usepackage{bm}

\usepackage{booktabs}
\usepackage{times}
\usepackage{rotating}
\usepackage{multirow}
\usepackage{amsmath,amsfonts,amssymb}
\newcolumntype{+}{>{\global\let\currentrowstyle\relax}}
\newcolumntype{^}{>{\currentrowstyle}}

\newcommand{\deltat}{\ensuremath{\Delta t}}

\DeclareMathAlphabet{\mathpzc}{OT1}{pzc}{m}{it}
\newcommand{\remove}[1]{}

\begin{document}

\title{Cliophysics: Socio-political Reliability Theory, Polity Duration and African Political (In)stabilities}

\author{Alhaji Cherif}
\affiliation{School of Human Evolution and Social Change, Arizona State University, Tempe, AZ, USA} 
\affiliation{Mathematical Institute, University of Oxford, 24-29 St. GilesÕ, Oxford OX1 3LB, UK}

\author{Kamal Barley}
\affiliation{School of Human Evolution and Social Change, Arizona State University, Tempe, AZ, USA} 

\begin{abstract}
\noindent Quantification of historical sociological processes have recently gained attention among theoreticians in the effort of providing a solid theoretical understanding of the behaviors and regularities present in sociopolitical dynamics. Here we present a reliability theory of polity processes with emphases on individual political dynamics of African countries. We found that the structural properties of polity failure rates successfully capture the risk of political vulnerability and instabilities in which $87.50\%$, $75\%$, $71.43\%$, and $0\%$ of the countries with monotonically increasing, unimodal, U-shaped and monotonically decreasing polity failure rates, respectively, have high level of state fragility indices. The quasi-U-shape relationship between average polity duration and regime types corroborates historical precedents and explains the stability of the autocracies and democracies.
\end{abstract}
\pacs{02.50.-r, 89.65.Cd, 89.65.Ef, 89.90.+n, 89.65.-s, 02.30.Zz}
\maketitle

 Beginning with the Berlin Conference in 1884 -- leading to the colonial despotism, to the National Conference of 1989 in Benin -- marking the beginning of the democratic renewal, Africa continues to experience various patterns of political-economic regulations, mostly shifting away from autocratic regimes and toward more open governance. The continent has seen an increasing number of incomplete transitions to democracy, resulting in a high vulnerability to outbreaks of conflict ~\cite{Auvinen,Bratton,Casper,Jaggers,Collier,Huntington}. Hypotheses for the proneness of the continent to conflicts include the inherent contradictions of being both partly open and repressive, the political incoherencies associated with anocratic (semi-democratic) regimes, economic underdevelopment, poor managerial styles, and ethnic politics and polarization ~\cite{Ellingsen,Ellingsen_etal,Francisco,Gurr1,Gurr2,Hauge,Heldt,Hibbs,Horowitz}. However, in the past decades, there has been a fewer political instabilities in anocratic regimes than predicted by historical data and previous studies. Here we develop a simple model of polity duration that explains historical data, and shows the risk of polity change depends on the structural properties of polity failure rates. Three cliophysical parameters ~\cite{Cliophysics} quantifying the properties of the risk of political vulnerabilities are estimated from the data, and are used to provide a scientific understanding of the underlying mechanisms of regime stability. We call this Òsocio-political reliability theory,Ó and it has relevant implications for policy evaluations and designs.
\\

Political processes within Africa have been characterized by waves of political instabilities, economic under-performances and other social factors with more than half of the sub-Saharan countries having experienced some forms of state-formation and/or post-independence civil conflicts since 1960, making the continent a testbed for exploring the determinants of political instabilities and state failures. As a result, we have seen an upsurge of literature dealing with the causes of civil wars and political unrests. Understanding the causes and consequences of these civil conflicts have been a focus of most of the recent social science research, particularly sociological, political and economic inquiries with in-depth studies highlighting the role of economic under-development and economic grievances in shaping risks of conflicts as illustrated by the works of Auvinen ~\cite{Auvinen}, Hauge and Ellingsen ~\cite{Hauge}, and Collier et al. ~\cite{Collier}; of monarchical tendency, oppression of minority groups and repressive managerial styles of african leaders as exemplified by the works of Francisco ~\cite{Francisco}, Lichbach ~\cite{Lichbach1}, and Moore ~\cite{Moore}; of the ethno-linguistic and religious diversity, ethnic fractionalization, political polarization, favoritism and nepotism as identified by the works of Ellingsen ~\cite{Ellingsen}, Horowitz~\cite{Horowitz}, Gurr ~\cite{Gurr1}, Lichbach ~\cite{Lichbach2} and Vanhanen ~\cite{Vanhanen1};  and comparatively, of the historical precedents in the works of Bratton and van de Walle ~\cite{Bratton}, Ellingsen and Gleditsch ~\cite{Ellingsen_etal}, Hibbs ~\cite{Hibbs}, Huntington ~\cite{Huntington}, and Muller and Weede ~\cite{Muller}. Some of these scholarly works have provided insights into the subtleties of socio-political processes underlying the historical precedents and constraints affecting the effectiveness of Africa's political behaviors, and have resulted in accumulation of empirical data and studies. Despite these accumulation of empirical knowledge,  theoretical contributions linking the polity dynamics, risk of political instabilities, and scientific understanding of the processes have been modest. In this paper, we provide quantitative evidence linking the structural properties of polity failure rates to variations of political dynamics, finding substantial supports and correlations between state fragility indices and the functional shape of polity failure rates. 

The model described herein links the levels of democracy, polity durations and properties associated with the risk of regime failure. To quantify the relationship between the polity duration and regime type, we use the polity score from the Polity IV data set ~\cite{Polity}, a scaled metric that measures regime governance and the extent to which countries are democratic. The score ranges between $-10$ for fully autocratic to $10$ for fully democratic, and was originally developed by Gurr and colleagues ~\cite{Gurr_etal,Polity}. For our purpose, we map the polity score from ~$(-10$, ~$10$) to ~$(0$, ~$1)$. In this paper, we use polity duration as a mean of assessing and measuring sociopolitical reliability, dynamics of which are affected by various micro- and macro-sociopolitical and economic processes. Possible processes that might affect polity dynamics are the changes in the governance and/or the structures of the government. This could be peaceful or rebellious transition (e.g., coup d' \' etats). Another example in which polity dynamics can change is in the managerial styles. That is, the adaptation of a larger policy changes and international pressures can influence the dynamics of polity. In other words, the government's attempts to minimize its probability of collapse or to maintain law and order can result in movements towards a repressive or an expansionary economic measure, and/or political repression or liberation. This in turn leads to changes in the level of democracy on the autocratic-democratic regime spectrum, and forces the state to move towards harshly autocratic or open governance. Therefore, polity change serves as an aggregative dynamical quantity that can provide some insights into the macro-sociopolitical behaviors originating from the complex interactions between the economic, social and political dynamics and domains. As a result, analyses presented herein focus on the dynamics of polity and its duration.

\begin{table}[ht]
\begin{center}
\begin{tabular}{lcclcrr}
\hline
\bf Regime Type &	& 	\bf Mean Polity duration, $\langle \deltat \rangle$  	&	& 	\bf 95\% CI 		&	&	 \bf N \\
\hline
Autocracies 	&	&	10.65									&	& 	(9.75, 11.56) 	&	&	164 \\
Anocracies	&	&	5.04										&	&	(5.02, 5.07)	&	&	53 \\
Democracies	&	&	10.20									&	& 	(8.42, 11.99)	&	&	78 \\
\hline
\end{tabular}
\caption{Empirical justification of average polity duration for different regime type from 1946-2008. rom the transformed polity score, autocracy is a polity with a score in the range [0, 1/3], anocracy within the range [1/3, 2/3], and democracy within the range [2/3, 1]}
\label{table:meanduration}
\end{center} 
\end{table}

\begin{figure}[hc]%
\begin{center}
\includegraphics[height = 6.6cm]{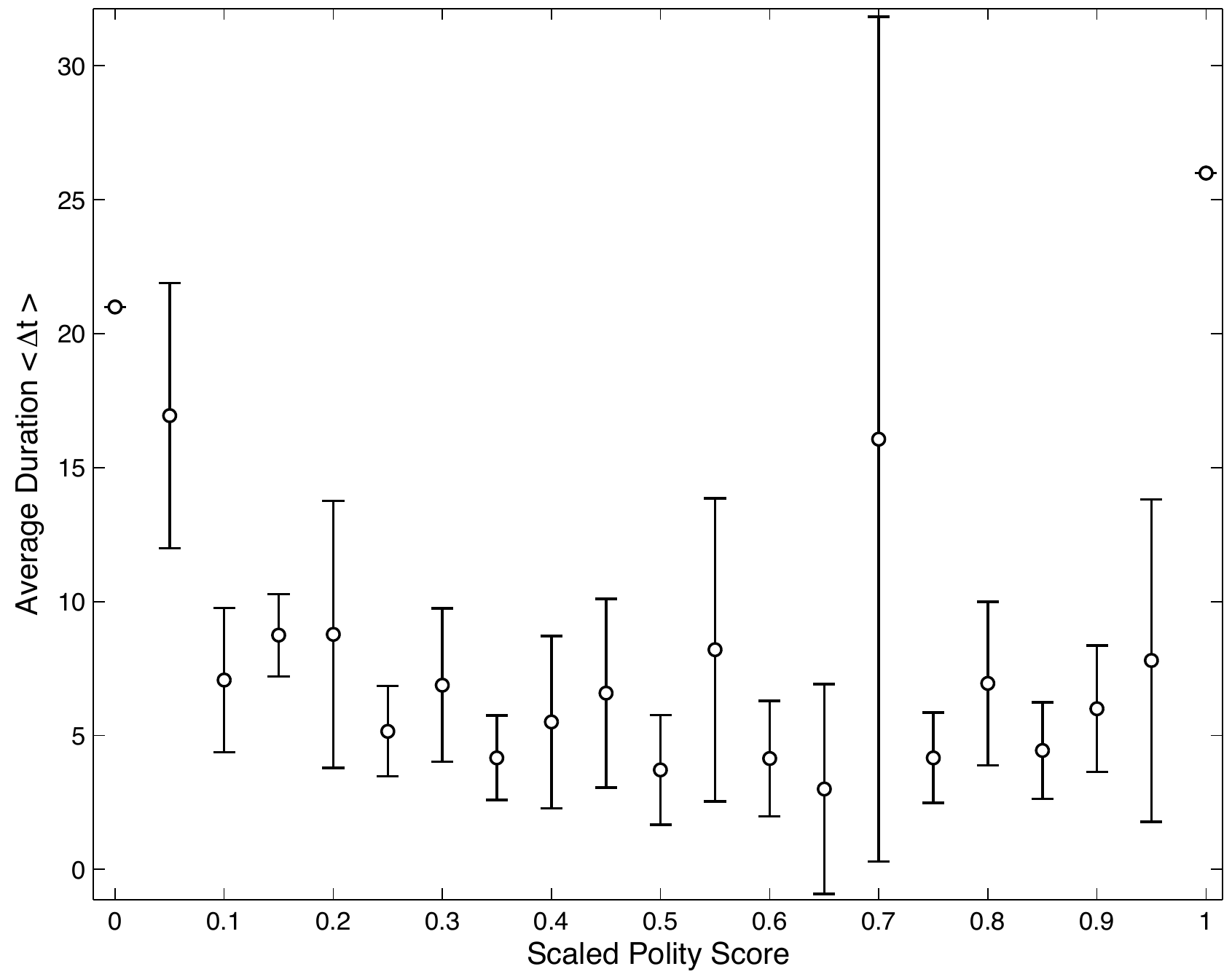}
\end{center}
\caption{{\bf Level of democracy and polity durations.} The figure shows the average polity duration, $\langle \deltat \rangle$ (years), with polities at either ends having longer duration}
\label{fig:meanduration}
\end{figure}
Aggregating polity levels, we found that the average polity durations for highly autocratic and institutionally consistent democratic regimes are longer than the anocratic ones (e.g., see Table ~\ref{table:meanduration}), in which autocracies  (e.g., scaled polity score range ~$= [0, 1/3])$ are estimated to be slightly longer than democracies in statistically insignificant way. However, fully or harshly autocracies (e.g., scaled polity score = 0) have shorter average polity duration than fully democracies (e.g., scaled polity score = 1). Table ~\ref{table:meanduration} summarizes our results. Figure ~\ref{fig:meanduration} shows an average duration as a function of scaled polity score, in which anocratic regimes are highly unstable with majority of the polities experiencing some regime change within the first five years. The regime volatility is inversely proportional to the average polity duration (e.g., $\mathcal{V}_{r} \propto \frac{1}{\langle \deltat \rangle}$). That is, the shorter the polity duration, the higher the political volatility and/or vulnerability to regime de-consolidation. Our findings corroborate the fact that both fully autocratic and democratic regimes are more durable and stable than anocratic regimes, and have been supported by numerous political scientists ~\cite{Krain,Levy,Lichbach1,Lichbach2,Vanhanen1,Vanhanen2,Ward_etal,Zanger}. However, we observed that anocratic regimes have relatively experienced a fewer political instabilities in the past decades than suggested by Table ~\ref{table:meanduration} and fig.~\ref{fig:meanduration}. To study the mechanisms that explain this, we look at the cliometric space of the shape parameters characterizing these polity failure rates to show that the structural properties of the polity failure rates capture the risks of political instabilities.

\begin{figure*}[ht]
\begin{center}
\includegraphics[scale=.5, height = 5.8cm]{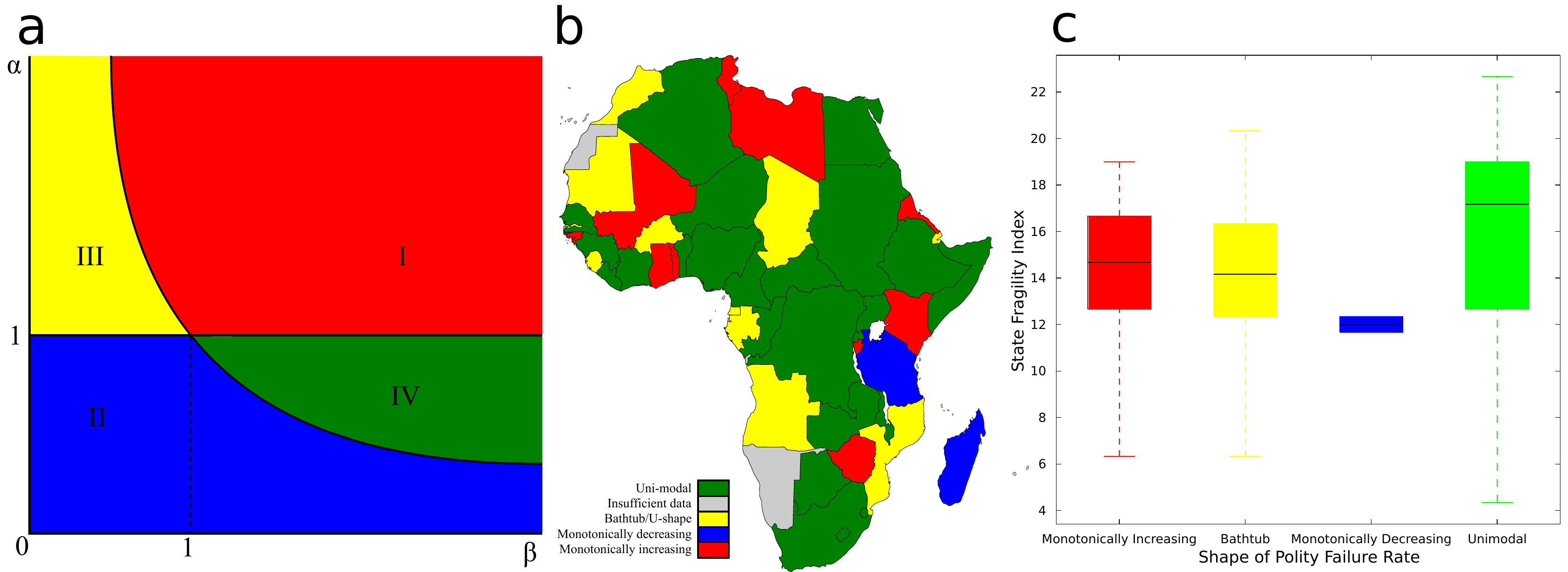}
\end{center}
\caption{{\bf Properties of the polity failure rate or political hazard.} The figures show: (a) functional type defining the risk of polity change (failure or hazard rate) in $\alpha-\beta$ plane separated by the curves $\alpha^{*} = 1$ and $\mathcal{C}^{*}_{p} = \sqrt{ \alpha \beta} = 1$; and (b) corresponding countries in Africa, as estimated by our model using least absolute regression with Trust-Region and Levenberg-Marquardt algorithms whenever possible. Interestingly, 87.50\%, 0\%, 71.43\%, and 75\% of African countries in regions I, II, III and IV, respectively, have extremely high state fragility indexes (c).
}
\label{fig:hazard}
\end{figure*}

To explore the direct role of functional shape of polity failure rate in explaining the historical risk of conflict and political instabilities, we use a modified Weibull function with a dispersive scale parameter, $\sigma > 0$, and ÒcliometricÓ shape parameters $\alpha > 0$ and $\beta > 0$ to characterize the distribution of polity duration, \deltat:
\begin{align}
S(\deltat; \sigma, \alpha, \beta) & \propto \left[ 1 - {\rm exp}\!\left( - \left[\frac{\deltat }{\sigma} \right]^{\alpha} \right) \right] ^{\beta} \enspace
\label{eq:model}
\end{align}
Fitting our model to Polity IV data set for forty-eight African countries that have sufficient data points for our estimation problem ~\cite{Gurr_etal,Polity,Collet,Cherif},  we can characterize the risk of polity change by four regions (e.g., see fig.~\ref{fig:hazard}$a$) in the cliometric diagram of shape parameters $\alpha$ and $\beta$ over which the failure rate as a function polity duration is, respectively, increasing (region I, e.g., $\mathcal{C}_{p} = \sqrt{\alpha \beta} \geq 1$ and $\alpha \geq 1$), decreasing (region II, e.g., $\mathcal{C}_{p}  \leq 1$ and $\alpha \leq 1$), U-shaped or bathtub-shaped (region III, e.g., $\mathcal{C}_{p}  < 1$ and $\alpha > 1$), and unimodal (region IV, e.g., $\mathcal{C}_{p} > 1$ and $\alpha < 1$). The Cliometric Number, $\mathcal{C}_{p} = \sqrt{\alpha \beta}$ is defined here as the geometric mean of the two shape parameters, and measures the initial growth of risk at the early stage of a new polity establishment. $\mathcal{C}_{p} < 1$ indicates an initial decay in the risk of polity change, while $\mathcal{C}_{p}$ provides an indication of a possible increase in the vulnerability of the polity. The cliometric shape parameter $\alpha$ determines the monotonicity. That is, the polity failure rate is monotone ($\alpha \geq 1$, $\mathcal{C}_{p} \geq 1$ or $\alpha  < 1$, $\mathcal{C}_{p} < 1$) or non-monotone ($\alpha > 1$, $\mathcal{C}_{p} < 1$ or $\alpha  > 1$, $\mathcal{C}_{p} < 1$) if the direction of both $\alpha$ and $\mathcal{C}_{p}$ are the same or opposite, respectively. The conditions above are based on the derivative of the failure rate, as defined by $h(\deltat; \sigma, \alpha, \beta) = \frac{S'(\deltat; \sigma, \alpha, \beta)}{1-S(\deltat; \sigma, \alpha, \beta)}$. That is, the analysis of $h'(\deltat; \sigma, \alpha, \beta)$ or equivalently of the transformation $\mathpzc{h}'(\sigma (\log z)^{1/\alpha}; \sigma, \alpha, \beta)$, where $z = exp((\deltat/\sigma)^{\alpha})$ determines the four distinct regions. Figure ~\ref{fig:hazard}$b$ show our empirical summary for African continent. Interestingly, we found that the majority of African countries have either U-shape (29.2\% or 14 countries) or uni-modal (50\% or 24 countries) polity failure rates. 

\begin{figure*}
\begin{center}
\includegraphics[width= 15cm, height = 20cm]{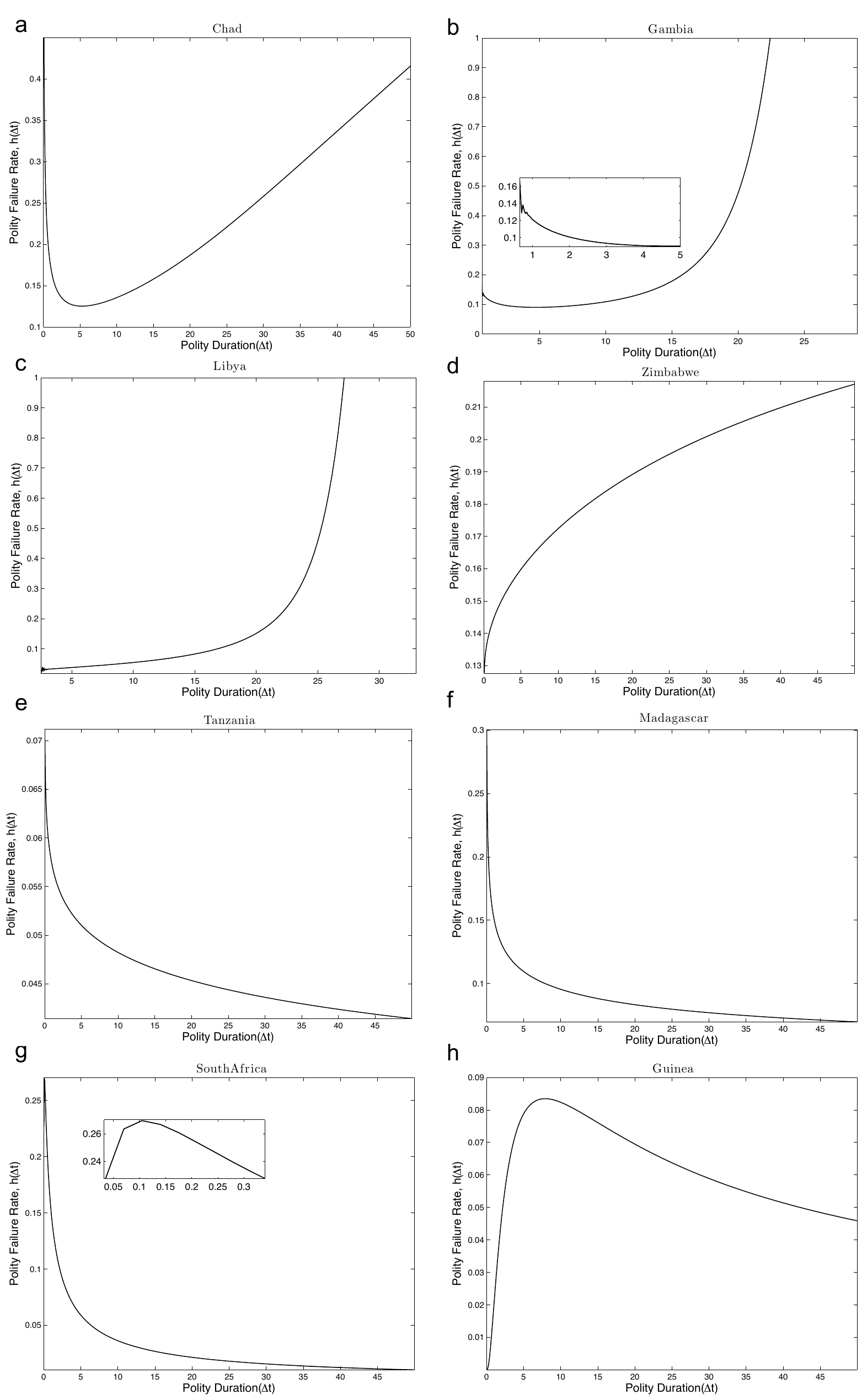}
\end{center}
\caption{{\bf Polity Failure Rates.} The figures show different properties for polity failure rates using estimated parameters and data sampled from eight different countries: (a) Chad with $\sigma_{c} = 14.81$, $\alpha_{c} = 1.946$, $\beta_{c} = 0.2602$; (b) Gambia with $\sigma_{g} = 20.9$, $\alpha_{g} = 10.22$, $\beta_{g} = 0.05641$; (c) Libya with $\sigma_{l} = 26.37$, $\alpha_{l} = 15.51$, $\beta_{l} = 0.07682$; (d) Zimbabwe with $\sigma_{z} = 7.151$, $\alpha_{z} = 1.133$, $\beta_{z} = 0.8861$; (e) Tanzania with $\sigma_{t} = 19.26$, $\alpha_{t} = 0.8893$, $\beta_{t} = 1.06$; (f) Madagascar with $\sigma_{m} = 8.15$, $\alpha_{m} = 0.8063$, $\beta_{m} = 0.9839$; (g) South Africa with $\sigma_{sa} = 3.11\times10^{-6}$, $\alpha_{sa} = 0.1042$, $\beta_{sa} = 71.07$; and (h) Guinea with $\sigma_{g} = 0.9487$, $\alpha_{g} = 0.4285$, $\beta_{g} = 11.04$, respectively. These parameters were estimated from empirical data using Least Absolute Regression method with Trust-Region and Levenberg-Marquardth algorithms, and were used to generate the polity failure rates: (a-b) Chad and Gambia have U-shaped or bathtub-shaped with Chad having a shorter constant failure rate and Gambia with shorter early life and longer constant polity failure rate before the increasing cumulative damage effect stage sets in; (c, d) Libya and Zimbabwe exhibit, respectively, convex and concave monotonically increasing failure rates; in contrast, Madagascar (e) and Tanzania (f) exhibit monotonically decreasing, and South Africa (g) and Guinea (h) have uni-modal polity failure rates, respectively. 
}
\label{fig:failure}
\end{figure*}

Bathtub-shaped or $U$-shaped polity failure rate suggests that there is an initial period of high level of uncertainty and civil unrest in a newly formed polity, followed by a long or short constant failure rate as dissidents abandoned or are integrated into the political system, and/or the government satisfies the political demands of the dissidents. However, as the loss of political legitimacy increases, the state effectiveness decreases, and the lifetime of the polity is prolonged, the risk of polity change is heightened due to cumulative risks. Depending on the regime type, especially for anocracies, this heightening of risk of polity change can easily lead to political instability and civil unrest. For autocracies, this implies the strengthening of repressive measures and institutions. Bathtub-shaped polity failure rate is reminiscent of three fundamental mechanisms driving the risk of polity change, namely the quality of governance, socio-political and economic factors, and management styles. We found that more than seventy percent (71.43\%) of the countries with bathtub-shape failure rates have extreme to serious level of state fragility index (e.g., see fig.~\ref{fig:hazard}$c$), a measure of state effectiveness and legitimacy in the areas of security, economics, development, and governance ~\cite{Polity}. 

In contrast to U-shaped polity failure rate, we also observe that other countries exhibit different functional properties in which the risk of polity change initially starts low and increases to a maximum, before decreasing with uni-modal failure rate. Twenty-four countries have failure rates exhibiting uni-modal behaviors. Interestingly, seventy-five percent (75\%) of the countries with this property have high to serious state fragility index. On the other hand, monotonically increasing and decreasing polity failure rates have fewer countries with these risk structures. Monotonically increasing polity hazard has the largest number of countries with more than eighty-seven percents (87.50\%) having high to serious state fragility index (e.g., see fig.~\ref{fig:hazard}$c$). In contrast, countries with the monotonically decreasing countries are the most stable with low state fragility index. Figure ~\ref{fig:failure} illustrates common features in the structural property of polity failure rates for eight different countries, irrespectively of the polity score. 

%
The picture that emerges from our socio-political reliability theory is that U-Shaped polity failure rate has both adaptation and damage cumulative phases. The size and the magnitude of each of these regions determine the nature of political instability. That is, two or more countries can have similar features of polity failure rates and still have different course of socio-political instability depending on either the magnitude of the risk (see figs.~\ref{fig:failure}$e$-$f$) and/or the size of certain region (e.g., longer or shorter constant failure region; see figs.~\ref{fig:failure}$a$-$b$) in the case of bathtub shaped, the distance between the peak of uni-modal failure rate and the y-axis (see figs. ~\ref{fig:failure}$g$-$h$), convexity and concavity of monotonically increasing and decreasing hazard rates (see figs.~\ref{fig:failure}$c$-$d$). For instance, Angola, Burkina Faso, Chad (e.g., see fig.~\ref{fig:failure}$a$), Comoros, Djibouti, Equatorial Guinea, Gabon, Gambia (e.g., see fig.~\ref{fig:failure}$b$), Mauritania, Morocco, Mozambique, and Sierra Leone are classified as having $U$-shaped risks of polity change. However, the socio-political environments (e.g., level of political factionalism), economic conditions, and structural details are different. Countries such as Burkina Faso, Comoros, Djibouti, Gabon, Gambia, Morocco, and Mozambique, for example, shorter left tail (decreasing failure rate) followed by a longer constant failure than other countries within this cliometric space of shape parameters ((see figs.~\ref{fig:failure}$a$-$b$)). These properties consistently capture most of the political behaviors in many of these countries. Gambia with longer constant polity failure rate, for example, had a democratic polity that lasted for forty years before becoming an autocracy. Burkina Faso and Djibouti on the other hand have started to carefully liberalize their regimes. Other features of polity failure rates exhibit similar sociopolitical interpretations supported by microscopic processes. For instance, the socio-political conditions in Zimbabwe (e.g., see fig.~\ref{fig:failure}$d$) continue to deteriorate as illustrated by monotonically increasing polity failure rates and as a result the risk of polity change continues to increase. On the other hand, because of the longer constant rate region in the case of Libya, the sociopolitical condition is relatively stable as compared to Zimbabwe (see figs.~\ref{fig:failure}$c$-$d$)). In contrast, We observed that countries with monotonically decreasing rates have lower rate of political instability. The political conditions in Madagascar and Tanzania have been relatively stable as captured by the monotone decrease in the polity failure rate (e.g., see figs.~\ref{fig:failure}$e$-$f$). Interestingly, South Africa has uni-modal polity failure rate. However, its functional signature is closely similar to that of monotonically decreasing failure rate with a steep initial increase. In the period leading to the unbanning of the ANC, the release of Nelson Mandela, and the repeal of apartheid policy, there was heightening risk of possible political unrests and changes leading to policies similar to the redistributive policies that were adopted by Zimbabwe after its independence. However, South Africa has been relatively stable since the apartheid era.

 
In conclusion, we find that the shift away from autocratic forms of governance in Africa provides strong evidence of the relationships between political violence and instability and democratization; and structural properties of the polity failure rates provide a mean to characterizing the risk of political instability in African countries, and other developing countries. Future research could illustrate whether greater understanding of the instantaneous behaviors of polity change coupled with regime type and other socio-political and economic factors can be used to design preventive and effective policies and/or mitigate political instability. An extension of our model using repairable system reliability theory with decision models may allow us to develop effective policies that minimize the cost of de-consolidating political institutions and the heightening risks of state failure, civil conflict and unrest.
\\
\newpage
\noindent {\bf Acknowledgments} The authors were supported in part by National Science Foundation through a Graduate Research Fellowship (to A.C.) and Louis Stokes Alliances for Minority Participation Fellowships (to A.C., K.B.), ASU Graduate College Doctoral Enrichment Fellowship (to A.C.), and Sloan Graduate Fellowships (to A.C).\\


\end{document}